# AN OPTIMIZED DISK SCHEDULING ALGORITHM WITH BAD-SECTOR MANAGEMENT

Amar Ranjan Dash[1], Sandipta Kumar Sahu[2] and B Kewal[3]

[1]Department of Computer Science, Berhampur University, Berhampur, India
[2]Department of Computer Science, NIST, Berhampur, India
[3]Department of Computer Science, Berhampur University, Berhampur, India

## ABSTRACT

*In high performance computing, researchers try to optimize the CPU Scheduling algorithms, for faster and efficient working of computers. But a process needs both CPU bound and I/O bound for completion of its execution. With modernization of computers the speed of processor, hard-disk, and I/O devices increases gradually. Still the data access speed of hard-disk is much less than the speed of the processor. So when processor receives a data from secondary memory it executes immediately and again it have to wait for receiving another data. So the slowness of the hard-disk becomes a bottleneck in the performance of processor. Researchers try to develop and optimize the traditional disk scheduling algorithms for faster data transfer to and from secondary data storage devices. In this paper we try to evolve an optimized scheduling algorithm by reducing the seek time, the rotational latency, and the data transfer time in runtime. This algorithm has the feature to manage the bad-sectors of the hard-disk. It also attempts to reduce power consumption and heat reduction by minimizing bad sector reading time.*

## 1. INTRODUCTION

In the era of high performance computing, most attention is on improving the capacity of computers by increasing their working speed. For that purpose, researchers try to optimize the performance measure of the traditional CPU scheduling algorithms [1], [2], [3] by reducing the average waiting time and turnaround time. Amar, Sandipta, and Sanjay [4] have developed an optimized CPU scheduling algorithm "DABRR" by introducing a concept of dynamic time quantum, made up by finding the mean of burst time of all process. In further study Amar et al. [5] selected specific priority features and proposed new scheduling algorithm "CSPDABRR". This algorithm made up of both round-robin and priority CPU scheduling algorithm. This algorithm execute the processes based on their priority and with that it also provides lesser average waiting time and average turnaround time then all pre-existing CPU scheduling algorithms. After arrival of multi core processor, the speed of processing becomes much faster with multi-threading technology. With evolution of computer, gradually the speed processor increases much more than the speed of hard-disk. All process needs both CPU time and I/O time for completion of its execution. For I/O operations, a process requests the operating system to access the disk to store or retrieve data. In multi-processing environment many processes are generated which in turn may generate multiple memory request, either to read some data from or to write some information to secondary storage device. So even with faster CPU processing it have to wait longer for retrieving data from secondary memory. So the slowness of the hard-disk becomes a bottleneck in performance of processor. That is why for controlling and providing the access to memory for all the processes operating system uses the concept of disk scheduling. Scheduling is a basic task of an operating system to schedule all computer resources before their use. To serve the read or write requests of multiple processes a disk should be efficient enough to give services to all processes and this can be achieved by various traditional disk scheduling algorithms [1], [2], [3]. Traditional disk scheduling algorithms are FCFS, SSTF, SCAN, C-SCAN, LOOK, and C-LOOK.

Track and sectors are basic conceptual parts based on which the disk scheduling algorithms serves all memory requests. Tracks are a set of concentric circular path on the surface of a disk or platter on which the data is retrieved or written on. On a track the data bits are stored in the form of magnetized bit pattern. Each track is further sub-divided into a number of partitions known as sectors. A sector stores a fixed amount of user accessible data.

The performance of a disk scheduling algorithm is measured in terms of it seek time, rotational latency and bandwidth. Seek time is the amount of time taken by the read/write head of a disk to move to a particular track where the information is present. Rotational latency is the amount of time taken by the read/write head of a hard disk to get to a specified sector in a track, by rotation of platter. It depends on the rotating speed of the disk. Data Transfer time is the summation of time required for selection of a read/write head and the time required to transfer the data. and number of bytes to be transferred. Bandwidth is the rate at which data is read from or written on hard disk. The main objective of scheduling algorithm is to process all memory request with lesser seek time and rotational latency. C. Mallikarjuna and P. Chitti Babu [6] have done a comparative analysis of all traditional disk scheduling algorithms based on their performance measure. Sukanya [7] have developed a simulation tool in C# to represent the working procedure of all disk scheduling algorithms with their statistics

As per the complexity the least complex algorithm is FCFS. As it is the simplest algorithm without any overload. But it requires more seek time other disk scheduling algorithm. Manish [8] proposed new disk scheduling algorithm IFCFS. This algorithm first completes all memory request in the path from initial head position to first memory request, then travel in reverse direction. But this algorithm fails in providing optimized seek time when the first memory request is not present either in starting or finishing end. Margo, Peter, John [9] proposed GSTF (Grouped Shortest Time First) and WSTF (Weighted Shortest Time First) disk scheduling mechanism to improve the performance of disk scheduling algorithm. In disk scheduling the main performance measure is seek time, rotational latency, and data transfer time. The LOOK algorithm provides the least seek time, among all traditional disk scheduling algorithms. So most of the researchers try to optimize the LOOK disk scheduling algorithm by reducing the seek time. Saman and Ritika [10] have proposed an algorithm, which calculates the difference between the highest request value and the lowest request value and then compares it with the current head position and then implements LOOK algorithm to service the requests. Sourav et al. in [11] and Sandipon et al. in [12] developed two similar disk scheduling algorithms.

These algorithms arrange the requests in ascending order and then calculates the left distance from current head position and then calculates the right distance from current head position and then decides the direction of head movement and implements LOOK algorithm to service the requests. Mahesh and Renuka [13] have developed a new disk scheduling algorithm "Sort Mid Current Comparison" (SMCC), to reduce the seek time. This algorithm arranges the requests in ascending order and then calculates the mid-point, then compares the mid-point with the current head position and decides the movement of head from initial head position and implements LOOK to service the requests. Jainil and Yash [14] have proposed a "Median Range Disk Scheduling Algorithm". This algorithm arranges the requests in ascending order and then finds the median range. After this it compares the current head position with the median range value.

If the head is between the median range, then SSTF algorithm is applied to service the requests or else LOOK algorithm is implemented.

Some of the researchers also try to reduce the rotational latency to optimize the disk scheduling algorithm. Priya and Supriya [15] have developed an algorithm to reduce the rotational latency of disk scheduling algorithm by appending the fuzzy logic. Kitae and Heonshik [16] have proposed two disk scheduling algorithms SRLF and SATF for decreasing the rotational latency and access time respectively. Karishma et al. [17] have developed a new disk scheduling algorithm "OTHDSA" (Optimized Two Headed Disk Scheduling Algorithm). This algorithm works on the hard-disk with two read/write heads for each side of platter. It works on the movement of Left and Right head, after measuring their distance from the position of memory request with lowest track and the position and memory request with high track respectively. Similarly, Avneesh, Abhijeet, and Abhishek [18] have developed another disk scheduling algorithm which will only work for hard disk which contains 3 head for each side of platter. Alexander [19] has performed a qualitative survey on all available disk scheduling algorithms, with a comparative analysis of STF & SPTF with all traditional scheduling algorithms.

Researchers try to analyse the amount of energy consumed and the amount of heat generated within hard disk. Anthony et al. [20] have analyzed the electrical energy consumed with in hard disk based on type of hardware and type of memory access. They also analyzed the energy consumed based on the position and size of data-chunk on hard disk. Wentao [21] have analyzed the energy consumed within hard disk and solid state drive. With complete analysis he states that energy consumed per accessing one bit is approximately 100 fj in hard disk and 0.35 fj in solid state drive.

All researchers just try to optimize the seek time and rotational latency based on current head position and memory request position based on Track and Sector. But no one considers platter and cylinder number for the optimization. They have also not proposed any algorithm for bad sector detection and management. The rest of the paper is structured as follows: in Section 2, we briefly discuss the traditional disk scheduling algorithms. In Section 3, we elaborate the proposed disk scheduling algorithm. In section 4 we made a comparative analysis of our algorithm with all traditional algorithms. Finally, in Section 5, we provide the concluding remarks and aspect of more improvement in future work.

## 2. TRADITIONAL ALGORITHMS

Due to the volatile nature of the CPU register, Cache, and Main Memory, the use of secondary storage devices such as Hard Disk came into existence. Hard disk is a secondary storage device which stores data on circular disks known as platters. All hardware technologies are evolving gradually to fulfil the requirement of high performance computing which is the basic need of modern computing. Still the speed of processor is much faster than the speed of hard-disk. To serve the read or write requests of multiple processes a disk should be efficient and fast.

All computer resources need to be scheduled, before use, for proper and faster utilization of them. This is basic work of the scheduling algorithms of operating system. In multi-processing environment many processes are generated which in turn generate many read or write request to store or retrieve data on a secondary storage device. But hard-disk can serve one memory request at a time. So all memory requests are stored in a queue. Two consecutive memory request may be far from each other (based on tracks), which so can result in greater disk arm movement. Hard drives are one of the slowest parts of computer system and thus need to be accessed in an efficient manner to match the speed of requests. To solve/manage above situation we need disk scheduling algorithm of operating system. The main objective of the disk scheduling algorithm is to provide the best seek time. The main goal of a disk scheduling algorithm is to reduce the starvation problem among the requests, to provide high throughput and to minimize the total amount of head movement. The performance of a hard disk can be enhanced by using various scheduling algorithm to give better seek time.

FCFS is the simplest of all disk scheduling algorithms. This algorithm uses first come first serve method which means the request which arrives first in the disk queue is serviced first. The main problem with this algorithm is that the head has to take large swings if the requests are far from each other which increases the seek time. In SSTF (shortest seek time first) the requests having the less seek time from the current head position is serviced first. In this algorithm the seek time of all requests are calculated in advance and are arranged in terms of their seek time in the disk queue. This algorithm causes starvation problem and an overhead is required to calculate the seek time.

SCAN algorithm also known as elevator algorithm. This algorithm starts the disk arm at one end of the disk and goes to till the other end servicing all the requests in between. After this the direction is reversed and the servicing continues in the reverse journey of the disk arm. The main disadvantage with this algorithm is that the requests have to wait long enough to be serviced and there is a lot of unnecessary head movement since the disk arm has to move till the ends even if there are no requests to be serviced. C-SCAN is known as circular scan. It is a modified version of scan. In this algorithm the disk arm starts from one end of a disk and moves to the other end servicing requests in between and then returns back from where it started without servicing any request and then the process continues. The main disadvantage with this algorithm is that it takes more head movement to service requests than SCAN.

LOOK algorithm services the requests in the same manner as that of SCAN but the difference is that instead of going till the end track (either higher or lower) it goes to the request which is nearest to the end track (either higher or lower). The main disadvantage is that we need to keep finding the nearest request in case of mass loading. C-LOOK is known as circular LOOK. This algorithm combines the features of C-SCAN and LOOK algorithm. In this algorithm the disk arm starts from its position and moves to a memory request nearer to one end of hand- disk, servicing requests in between. Then returns back from there to the memory request nearer to other end and then the process continues.

At present the platters of hard-disk are made up of aluminium, to make the hard-disk as light weight as possible. While manufacturing the both side platters are coated with a vacuum deposition process called magnetron sputtering of magnetic substances, over which a thin film of dust carbon power is over coated by sputtering process. So a nanometre thin layer of polymeric lubricant layer gets deposited on the top of surface of platter by doping all platters into lubricant solvent solution. Then the disk is buffered by various process to eliminate small defects by typical sensor detection techniques, radially a bad sector or bit pattern is repaired by defuse dusting of carbon compounds. The surface of typical hard disks was coated with diamagnetic oxide or paramagnetic material. The surface of platters is buffered to ensure the error correction purposes which can store a billion bits per square inches.

The Platters must have to kept dust free during development of hard-disk. To eliminate internal contamination from dust, air pressure is equalized by vacuum filters and platters are hermetically placed in the disk case by a partial vacuum field with reading writing head in prior position, this is called Hard Disk Assembly. Traditionally, the read/write heads use infrared and blue ray technology for reading the data stored in magnetic bit pattern. But recently companies use LED with laser for that same purpose. Self-monitoring, Analysis and Reporting Technology (S.M.A.R.T) is the common technology used by all hard drives as in Serial DATA or Parallel DATA drives for analysis and error detection. ECC (error correction code), (LDPC) Low Density Parity Checking Code are used till now as error correction code by Modern TOSIBHA, SEGATE, SAMSUNG Drives of from capacity of 120 GB.

## 3. OUR PROPOSAL

Disk scheduling algorithm is the specific task of operating system to serve all memory requests with as less time as possible. Hard disks typically have a number of platters of same radius mounted on a spindle. A platter can store information on both sides. Each platter of a hard disk is divided into concentric circles known as 'Tracks'. Track is a circular path on the surface of a disk or platter on which the data is stored. On a track the data bits are stored in the form of magnetized bit pattern. The number of tracks on the single surface of a drive is exactly equal to the number of cylinders of the drive. Each track on a platter is sub-divided into number of virtual parts known as sectors. Each sector holds a fixed amount of data. Traditionally a sector holds 512 bytes for a hard disk. Platter, Track and sectors are basic parts based on which the disk scheduling algorithms serves all memory requests. Disk queue is a queue in which all the I/O requests which are to be serviced are stored.

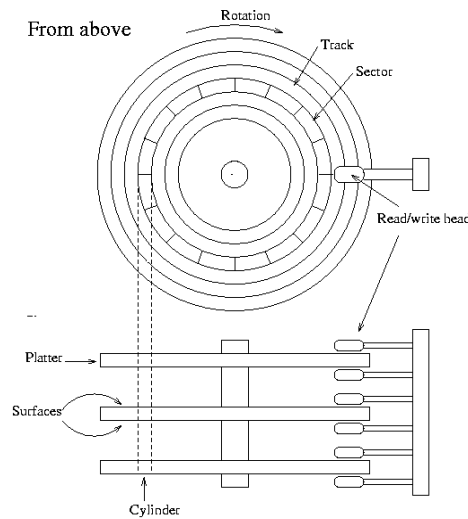

Figure 1: Top and Side View of Hard-disk

The performance of a disk scheduling algorithm is measured in terms of it disk access time which again depends upon seek time, rotational latency and bandwidth. Seek time is the amount of time taken by the read/write head of a disk to move to a desired track where the recent memory request is present. After reaching a particular track the read/write head have to wait, till it reaches a particular sector by rotation of platter, to serve a request. Rotational latency is the amount of time taken by the read/write head of a hard disk to get to a specified sector in a track. It is also known as rotational delay. Bandwidth is the rate at which data is read from or written on hard disk. Transfer time is the time to transfer the data. Disk access time is the sum of the seek time, rotational latency and transfer time. Similarly, the Disk response time is the combination of disk access time and disk querying time. The main objective of disk scheduling algorithm is to process all memory requests with lesser Disk Access Time. Figure 1 Represents the top and side view of Hard disk.

All Traditional algorithms only try to minimize the seek time during the process of servicing the memory requests. In all referred papers, researchers try to reduce the seek time and rotational latency for optimizing the disk scheduling algorithm. Those algorithms are perfectly applicable for hard-disk with single platter. But in reality, all hard disks are made up of multiple platters.
We consider Platter, Track, Cylinder, and Sector number as primary input for memory request. Table 1 show the proposed structure of disk queue for our algorithm. This disk queue is made up of Linked-List. In this algorithm we proposed a mechanism to detect and manage the bad sectors of a hard disk, by creation and maintenance of an additional Linked List termed as "Bad Sector List", as shown in Table 2.

Table1: Structure of Disk Queue

| Platter | Track | Sector | Index | R/W | BSI |
|---------|-------|--------|-------|-----|-----|
|         |       |        |       |     |     |

Table 2: Structure of Bad Sector List

| Index | BSI | Temp/Permanent | Prescribed Bit | Finalised |
|-------|-----|----------------|----------------|-----------|
|       |     |                |                |           |

### 3.1 Proposed Algorithm

Here we proposed a new optimized disk scheduling algorithm named as MODSSM (Modern Optimized Disk Scheduling with Bad-Sector Management).

**MODSBSM:**
  IHP: Initial Head Position
  IHPt: Track of Initial Head Position
  IHPp: Platter of Initial Head Position
  IHPs: Sector of Initial Head Position
  CHP: Current Head Position
  CHPt: Track of Current Head Position
  CHPp: Platter of Current Head Position
  CHPs: Sector of Current Head Position
  LD: Left Distance
  RD: Right Distance
  Nr: Number of memory request
  NP: Number of Platter
  BSQL: Bad Sector Queue List
  TSKT: Total Seek Time
  TRL: Total Rotational Latency
  TDTT: Total Data Transfer Time
  TDAT: Total Disk Access Time
  i1, i2: Different Indexes
  BSI: Bad Sector Index
  Index: a number which holds platter, track, and sector details of a memory block within hard-disk
  BSM(): A function for management of detected bad sector

Step 1:    Insert all memory request in partition list of hard disk.
Step 2:    TSKT = 0
Step 3:    TRL = 0
Step 4:    TDTT = 0
Step 5:    Rearrange the partition list, in ascending order of Track,
Step 6:    IHP: initial disk head position
Step 7:    LD = (IHPt – track value of first node of partition table)
Step 8:    RD = (track value of last node of partition table– IHPt)

```
Step 9:     //Decide the direction of head movement from IHP
Step 10:    If (LD < RD)

Step 11:      Rearrange the partition list,
Step 12:        within each track, arrange the sectors in ascending order.
Step 13:        Then within each sector, arrange the platters in ascending order.
Step 14:      i1 = 0
Step 15:      SKT = SKT+ LD
Step 16:      //For track-wise traversing of the partition list
Step 17:      Repeat Step 18 to 56 till (i1 < Nr)
Step 18:        Move the Head to the track section of i1 node
Step 19:        i2=i1
Step 20:        //Complete all memory request from different platters & sectors belongs to the i2 cylinder
Step 21:        Repeat Step 22 to 51 till (Track of i2 == Track of i1)
Step 22:          If (r/w head is unable to read a bit)
Step 23:          //bad sector means unreadable and non-writable
Step 24:            If (BSI == 2)
Step 25:              BSM(index);
Step 26:            Else If (BSI == 0)
Step 27:              BSI++
Step 28:              Delete the block, from that position of the partition list & add at end of the partition list
Step 29:            Else
Step 30:              BSI++
Step 31:              Send the node to the Bad-Sector Table
Step 32:          Else
Step 33:            If(i2==0)
Step 34:              If (IHPs == sector of i2)
Step 35:                TRL = TRL + 0
Step 36:              Else If (sector of i2 > IHPs)
Step 37:                TRL = TRL + (sector of i2-IHPs)
Step 38:              Else
Step 39:                TRL =TRL + (7-IHPs+1+sector of i2)
Step 40:              TDTT = TDTT + (Absolute (Platter of i2-Platter of IHPp))
Step 41:              Complete the Memory request of i2 node
Step 42:            Else
Step 43:              If (sector of (i2-1) == sector of i2)
Step 44:                TRL = TRL + 0
Step 45:              Else If (sector of i2 > sector of (i2-1))
Step 46:                TRL = TRL + (sector of i2-sector of (i2-1))
Step 47:              Else
Step 48:                TRL =TRL + (7-sector of (i2-1) +1+sector of i2)
Step 49:              TDTT = TDTT + (Absolute (Platter of i2-Platter of (i2-1)) )
Step 50:              Complete the Memory request of i2 node
Step 51:          i2++
Step 52:        if (i2 < Nr)
Step 53:          SKT = SKT + (Track of i2 – Track of i1)
Step 54:          i1=i2
Step 55:        else
Step 56:          goto step 109
Step 57:    Else If (RD < LD)
Step 58:      Rearrange the partition list,
Step 59:        within each track, arrange the sectors in descending order.
Step 60:        Then within each sector, arrange the platters in descending order.
Step 61:      i1 = Nr-1
Step 62:      SKT = SKT+ RD
Step 63:      //For track-wise traversing of the partition list
Step 64:      Repeat Step 65 to 103 till (i1 >= 0)
Step 65:        Move the Head to the track section of i1 node
Step 66:        i2=i1
Step 67:        //Complete all memory request from different platters & sectors belongs to the track (or cylinder) of i1
Step 68:        Repeat Step 69 to 98 till (Track of i2 == Track of i1)
Step 69:          If (r/w head is unable to read a bit)
Step 70:          //bad sector means unreadable and non-writable
Step 71:            If (BSI == 2)
Step 72:              BSM(index);
Step 73:            Else If (BSI == 0)
Step 74:              BSI++
Step 75:              Delete the block, from that position of the partition list & add at end of the partition list
Step 76:            Else
Step 77:              BSI++
Step 78:              Send the node to the Bad-Sector Table
```

```
Step 79:      Else
Step 80:        If(i2==Nr-1)
Step 81:         If (IHPs == sector of i2)
Step 82:           TRL = TRL + 0
Step 83:         Else If (sector of i2 > IHPs)
Step 84:           TRL = TRL + (sector of i2-IHPs)
Step 85:         Else
Step 86:           TRL =TRL + (7-IHPs+1+sector of i2)
Step 87:         TDTT = TDTT + (Absolute (Platter of i2-Platter of IHPp))
Step 88:         Complete the Memory request of i2 node
Step 89:        Else
Step 90:         If (sector of (i2+1) == sector of i2)
Step 91:           TRL = TRL + 0
Step 92:         Else If (sector of i2 > sector of (i2+1))
Step 93:           TRL = TRL + (sector of i2-sector of (i2+1))
Step 94:         Else
Step 95:           TRL =TRL + (7-sector of (i2+1) +1+sector of i2)
Step 96:         TDTT = TDTT + (Absolute (Platter of i2-Platter of (i2+1)))
Step 97:         Complete the Memory request of i2 node
Step 98:      i2—
Step 99:      if (i2 >= 0)
Step 100:       SKT = SKT + (Track of i1 – Track of i2)
Step 101:       i1=i2
Step 102:     else
Step 103:       goto step 109
Step 104:  Else
Step 105:    If (Head is recently moving from Higher Track to Lower Track)
Step 106:      goto Step 11
Step 107:    Else
Step 108:      goto Step 58
Step 109:  If (partition list is not empty)
Step 110:    Goto Step 5
Step 111:  Else
Step 112:    TDAT = TSKT + TRL + TDTT
Step 113:    End of algorithm
```

**BSM(Index):**
Finalized: a column of Bad Sector list table to check whether any value is finalized for that particular bad sector

```
Step 1:   if (finalized ==1)
Step 2:     Perform memory operation
Step 3:   Else
Step 4:     If (Prescribed Bit is applicable)
Step 5:       Perform memory operation
Step 6:     Else
Step 7:       Change the preferred-bit
Step 8:       Perform memory operation
```

### 3.2 Illustration

In this section we have analyzed the execution of the proposed algorithm. All memory request generated by the processes are stored in disk queue. The Disk queue is made up of Double-Linked-List. Here all records of disk queue, as shown in Table 1, are individual nodes of disk queue list. All nodes have eight sections. Six middle section for holding information of six columns. The first and last section of all nodes used to store the address of its previous and next node, for link. First all memory request nodes are arranged based on track.

Then check whether the initial position of the r/w head is nearer to track of first node of disk queue list or nearer to track of last node of disk queue list. If the initial position of r/w head is nearer to the first node of disk queue list, first arrange the sectors (with similar tracks) in ascending order, then with in each tack-sector combination arrange the platters in ascending order, then finally serve all memory requests from first to last node of disk queue list. Or if the initial position of r/w head is nearer to the last node of disk queue list, first arrange the sectors (with similar tracks) in descending order, then with in each tack-sector combination arrange the platters in descending order, then finally serve all memory requests from last to first node of disk queue list. Figure 2 represents the internal structure of a hard-disk.

Seek-time, Rotational-Latency, and data transfer time decide the disk access time, which is the basic performance measure of Disk scheduling algorithm. Seek-time depends on the speed of head movement. Rotational Latency depends on the revolution speed of platter. Similarly, the data transfer time depends on the disk bandwidth of the hard-disk. Out of these three the slowest one is the movement of r/w head. So most stress have been given to reduce the seek time. So once the head reach to a particular track, the algorithm completes all memory requests from that cylinder, irrespective of different platter or sector.

All Platters are connected through spindle. Similarly, all r/w heads are connected to a single actuator arm. So all r/w heads are move together simultaneously and remain at same track number, irrespective of any platter. In other words, all r/w heads are present in single cylinder. Once r/w head reach a particular track, at first it read all memory request present in different platters of same sector. After that it performs all memory request present in different platters of next sector. Similar process continuous till the algorithm processed all memory request of same cylinder, till the last platter.

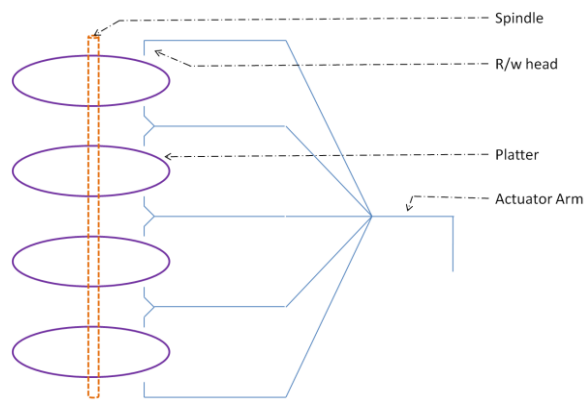

Figure 2**:** Internal structure of hard-disk

During the process of accessing data, if concerned bit is accessible, then normal memory operation occurs. But if it is not accessible, then that bit is called as bad sector. In that case the proposed algorithm increases the bad sector index by 1. Then the proposed algorithm removes the respective row from the disk queue and add it at the rear end of the disk queue. If one bit detected twice as bad sector, then that bit's information is transferred to bad sector table. Hard disk stores all information in binary bit format. In bad sector table there is a column which stored a preferable bit for each bad sector. At first the preferable bit of all bad sectors is considered as 'zero'. During execution of a bad sector, for the first time, algorithm check if the concerned memory request gets the perfect output by considering the preferred it as "Zero". If its output is perfect, then the finalized column of that index become one and after that for all read operation regarding that index, the algorithm processed the memory operation by considering the preferred bit as "zero". If its output is not perfect, then the algorithm converts the finalized column of that index into one, after inversing the preferred it cell of that memory index. After that for all read operation regarding that memory index, the algorithm processed the memory operation by considering the preferred bit as one.

## 4. COMPARATIVE ANALYSIS

### 4.1 Assumption

First we divide the problems into two types based on number of platter present in concerned hard-disk (hard-disk with single platter and hard-disk with multi platter). We further divide each into 3 more types based on the track number of memory requests (in ascending order, descending order, & random order). We analyzed all disk algorithms based on six situations. In each we have considered twenty independent memory requests with their particular memory request index. Corresponding Platter, Track, and Sector number of all memory requests are known before execution. The time required for sorting of memory requests assumed as zero. In each case all memory requests are processed by 6 traditional (FCFS, SSTF, SCAN, C-SCAN, LOOK, C-LOOK), 5 derived (ODSA, HDSA, RP-10, SMCC, MRSA) and proposed (MODSBSM) disk scheduling algorithm. Unlike others Saman Rasool has not given any specific name to his proposed disk scheduling algorithm. So we represent that algorithm as RP-10, as we referred that paper in 10$^{th}$ position in reference list. Seek Time is calculated by finding out the head movement. Rotational Latency is found by measuring amount of rotation needed. Data Transfer time is calculated by adding the time required for selection of particular read/write head and the time required for data transfer (assumed as one). Disk Access Time is the summation of Seek Time, Rotational Latency, and Data Transfer Time. At first booting of a operating system the BSI of all memory index is assumed as 0.

### 4.2 Case-1

Here we considered a situation with all memory request in ascending order in Single Platter Hard-disk. Figure 3 represents the memory request of Case-1. Initially the head is positioned at 65t1p4s Index. Figure 4 represents the working of traditional, referred, and proposed disk scheduling algorithm to access all Memory Requests of Case-1. Figure 5 represents the total seek time and disk access time required with respect to each disk scheduling algorithms for complete all memory requests as per Case-1.

| Platter | 1 | 1 | 1 | 1 | 1 | 1 | 1 | 1 | 1 | 1 | 1 | 1 | 1 | 1 | 1 | 1 | 1 | 1 | 1 | 1 |
|---|---|---|---|---|---|---|---|---|---|---|---|---|---|---|---|---|---|---|---|---|
| Track | 15 | 48 | 48 | 48 | 48 | 60 | 90 | 90 | 90 | 108 | 108 | 108 | 126 | 168 | 168 | 168 | 179 | 179 | 179 | 196 |
| Sector | 2 | 7 | 0 | 4 | 6 | 1 | 6 | 1 | 4 | 7 | 2 | 5 | 7 | 1 | 5 | 4 | 4 | 2 | 6 | 7 |
| Index | 15t1p2s | 48t1p7s | 48t1p0s | 48t1p4s | 48t1p6s | 60t1p1s | 90t1p6s | 90t1p1s | 90t1p4s | 108t1p7s | 108t1p2s | 108t1p5s | 126t1p7s | 168t1p1s | 168t1p5s | 168t1p4s | 179t1p4s | 179t1p2s | 179t1p6s | 196t1p7s |
| R/w | | | | | | | | | | | | | | | | | | | | |
| BSI | 0 | 0 | 0 | 0 | 0 | 0 | 0 | 0 | 0 | 0 | 0 | 0 | 0 | 0 | 0 | 0 | 0 | 0 | 0 | 0 |

Figure 3: Disk queue for memory request of case-1

| | FCFS | | | | | SSTF | | | | | SCAN | | | | | C-SCAN | | | | | LOOK | | | | | C-LOOK | | | | |
|---|---|---|---|---|---|---|---|---|---|---|---|---|---|---|---|---|---|---|---|---|---|---|---|---|---|---|---|---|---|---|
| T S P | ST | RL | DTT | DAT | T S P | ST | RL | DTT | DAT | T S P | ST | RL | DTT | DAT | T S P | ST | RL | DTT | DAT | T S P | ST | RL | DTT | DAT | T S P | ST | RL | DTT | DAT |
| 65 4 1 | | | | | 65 4 1 | | | | | 65 4 1 | | | | | 65 4 1 | | | | | 65 4 1 | | | | | 65 4 1 | | | | |
| 15 2 1 | 50 | 6 | 1 | 57 | 60 1 1 | 5 | 5 | 1 | 11 | 60 1 1 | 5 | 5 | 1 | 11 | 60 1 1 | 5 | 5 | 1 | 11 | 60 1 1 | 5 | 5 | 1 | 11 | 60 1 1 | 5 | 5 | 1 | 11 |
| 48 7 1 | 33 | 5 | 1 | 39 | 48 6 1 | 12 | 5 | 1 | 18 | 48 6 1 | 12 | 5 | 1 | 18 | 48 6 1 | 12 | 5 | 1 | 18 | 48 6 1 | 12 | 5 | 1 | 18 | 48 6 1 | 12 | 5 | 1 | 18 |
| 48 0 1 | 0 | 1 | 1 | 2 | 48 4 1 | 0 | 6 | 1 | 7 | 48 4 1 | 0 | 6 | 1 | 7 | 48 4 1 | 0 | 6 | 1 | 7 | 48 4 1 | 0 | 6 | 1 | 7 | 48 4 1 | 0 | 6 | 1 | 7 |
| 48 4 1 | 0 | 4 | 1 | 5 | 48 0 1 | 0 | 4 | 1 | 5 | 48 0 1 | 0 | 4 | 1 | 5 | 48 0 1 | 0 | 4 | 1 | 5 | 48 0 1 | 0 | 4 | 1 | 5 | 48 0 1 | 0 | 4 | 1 | 5 |
| 48 6 1 | 0 | 2 | 1 | 3 | 48 7 1 | 0 | 7 | 1 | 8 | 48 7 1 | 0 | 7 | 1 | 8 | 48 7 1 | 0 | 7 | 1 | 8 | 48 7 1 | 0 | 7 | 1 | 8 | 48 7 1 | 0 | 7 | 1 | 8 |
| 60 1 1 | 12 | 3 | 1 | 16 | 15 2 1 | 33 | 3 | 1 | 37 | 15 2 1 | 33 | 3 | 1 | 37 | 15 2 1 | 33 | 3 | 1 | 37 | 15 2 1 | 33 | 3 | 1 | 37 | 15 2 1 | 33 | 3 | 1 | 37 |
| 90 6 1 | 30 | 5 | 1 | 36 | 90 6 1 | 75 | 4 | 1 | 80 | 0 | | 15 | | | 0 | | 15 | | | 90 6 1 | 75 | 4 | 1 | 80 | 196 7 1 | 181 | 5 | 1 | 187 |
| 90 1 1 | 0 | 3 | 1 | 4 | 90 1 1 | 0 | 3 | 1 | 4 | 90 6 1 | 90 | 4 | 1 | 95 | 199 | | | | 199 | 90 1 1 | 0 | 3 | 1 | 4 | 179 6 1 | 17 | 7 | 1 | 25 |
| 90 4 1 | 0 | 3 | 1 | 4 | 90 4 1 | 0 | 3 | 1 | 4 | 90 1 1 | 0 | 3 | 1 | 4 | 196 7 1 | 3 | 5 | 1 | 9 | 90 4 1 | 0 | 3 | 1 | 4 | 179 2 1 | 0 | 4 | 1 | 5 |
| 108 7 1 | 18 | 3 | 1 | 22 | 108 7 1 | 18 | 3 | 1 | 22 | 90 4 1 | 0 | 3 | 1 | 4 | 179 6 1 | 17 | 7 | 1 | 25 | 108 7 1 | 18 | 3 | 1 | 22 | 179 4 1 | 0 | 2 | 1 | 3 |
| 108 2 1 | 0 | 3 | 1 | 4 | 108 2 1 | 0 | 3 | 1 | 4 | 108 7 1 | 18 | 3 | 1 | 22 | 179 2 1 | 0 | 4 | 1 | 5 | 108 2 1 | 0 | 3 | 1 | 4 | 168 4 1 | 11 | 0 | 1 | 12 |
| 108 5 1 | 0 | 3 | 1 | 4 | 108 5 1 | 0 | 3 | 1 | 4 | 108 2 1 | 0 | 3 | 1 | 4 | 179 4 1 | 0 | 2 | 1 | 3 | 108 5 1 | 0 | 3 | 1 | 4 | 168 5 1 | 0 | 1 | 1 | 2 |
| 126 7 1 | 18 | 2 | 1 | 21 | 126 7 1 | 18 | 2 | 1 | 21 | 108 5 1 | 0 | 3 | 1 | 4 | 168 4 1 | 11 | 0 | 1 | 12 | 126 7 1 | 18 | 2 | 1 | 21 | 168 1 1 | 0 | 4 | 1 | 5 |
| 168 1 1 | 42 | 2 | 1 | 45 | 168 1 1 | 42 | 2 | 1 | 45 | 126 7 1 | 18 | 2 | 1 | 21 | 168 5 1 | 0 | 1 | 1 | 2 | 168 1 1 | 42 | 2 | 1 | 45 | 126 7 1 | 42 | 6 | 1 | 49 |
| 168 5 1 | 0 | 4 | 1 | 5 | 168 5 1 | 0 | 4 | 1 | 5 | 168 1 1 | 42 | 2 | 1 | 45 | 168 1 1 | 0 | 4 | 1 | 5 | 168 5 1 | 0 | 4 | 1 | 5 | 108 5 1 | 18 | 6 | 1 | 25 |
| 168 4 1 | 0 | 7 | 1 | 8 | 168 4 1 | 0 | 7 | 1 | 8 | 168 5 1 | 0 | 4 | 1 | 5 | 126 7 1 | 42 | 6 | 1 | 49 | 168 4 1 | 0 | 7 | 1 | 8 | 108 2 1 | 0 | 5 | 1 | 6 |
| 179 4 1 | 11 | 0 | 1 | 12 | 179 4 1 | 11 | 0 | 1 | 12 | 168 4 1 | 0 | 7 | 1 | 8 | 108 5 1 | 18 | 6 | 1 | 25 | 179 4 1 | 11 | 0 | 1 | 12 | 108 7 1 | 0 | 5 | 1 | 6 |
| 179 2 1 | 0 | 6 | 1 | 7 | 179 2 1 | 0 | 6 | 1 | 7 | 179 4 1 | 11 | 0 | 1 | 12 | 108 2 1 | 0 | 5 | 1 | 6 | 179 2 1 | 0 | 6 | 1 | 7 | 90 4 1 | 18 | 5 | 1 | 24 |
| 179 6 1 | 0 | 4 | 1 | 5 | 179 6 1 | 0 | 4 | 1 | 5 | 179 2 1 | 0 | 6 | 1 | 7 | 108 7 1 | 0 | 5 | 1 | 6 | 179 6 1 | 0 | 4 | 1 | 5 | 90 1 1 | 0 | 5 | 1 | 6 |
| 196 7 1 | 17 | 1 | 1 | 19 | 196 7 1 | 17 | 1 | 1 | 19 | 179 6 1 | 0 | 4 | 1 | 5 | 90 4 1 | 18 | 5 | 1 | 24 | 196 7 1 | 17 | 1 | 1 | 19 | 90 6 1 | 0 | 5 | 1 | 6 |
| Total | 231 | 67 | 20 | 318 | Total | 231 | 75 | 20 | 326 | 196 7 1 | 17 | 1 | 1 | 19 | 90 1 1 | 0 | 5 | 1 | 6 | Total | 231 | 75 | 20 | 326 | Total | 337 | 90 | 20 | 447 |
| Avg | 11.55 | 3.35 | 1.00 | 15.90 | Avg | 11.55 | 3.75 | 1.00 | 16.30 | Total | 261 | 75 | 20 | 356 | 90 6 1 | 0 | 5 | 1 | 6 | Avg | 11.55 | 3.75 | 1.00 | 16.30 | Avg | 16.85 | 4.50 | 1.00 | 22.35 |
| | | | | | | | | | | Avg | 13.05 | 3.75 | 1.00 | 17.80 | Total | 373 | 90 | 20 | 483 | | | | | | | | | | |
| | | | | | | | | | | | | | | | Avg | 18.65 | 4.50 | 1.00 | 24.15 | | | | | | | | | | |

| | ODSA | | | | | HDSA | | | | | RP-10 | | | | | SMCC | | | | | MRSA | | | | | MODSBSM | | | | |
|---|---|---|---|---|---|---|---|---|---|---|---|---|---|---|---|---|---|---|---|---|---|---|---|---|---|---|---|---|---|---|
| T P S | ST | RL | DTT | DAT | T S P | ST | RL | DTT | DAT | T S P | ST | RL | DTT | DAT | T S P | ST | RL | DTT | DAT | T S P | ST | RL | DTT | DAT | T S P | ST | RL | DTT | DAT |
| 65 4 1 | | | | | 65 4 1 | | | | | 65 4 1 | | | | | 65 4 1 | | | | | 65 4 1 | | | | | 65 4 1 | | | | |
| 15 2 1 | 50 | 6 | 1 | 57 | 60 1 1 | 5 | 5 | 1 | 11 | 90 6 1 | 25 | 2 | 1 | 28 | 15 2 1 | 50 | 6 | 1 | 57 | 15 2 1 | 50 | 6 | 1 | 57 | 15 2 1 | 50 | 6 | 1 | 57 |
| 48 7 1 | 33 | 5 | 1 | 39 | 48 6 1 | 12 | 5 | 1 | 18 | 90 1 1 | 0 | 3 | 1 | 4 | 48 7 1 | 33 | 5 | 1 | 39 | 48 7 1 | 33 | 5 | 1 | 39 | 48 0 1 | 33 | 6 | 1 | 40 |
| 48 0 1 | 0 | 1 | 1 | 2 | 48 4 1 | 0 | 6 | 1 | 7 | 90 4 1 | 0 | 3 | 1 | 4 | 48 0 1 | 0 | 1 | 1 | 2 | 48 0 1 | 0 | 1 | 1 | 2 | 48 4 1 | 0 | 4 | 1 | 5 |
| 48 4 1 | 0 | 4 | 1 | 5 | 48 0 1 | 0 | 4 | 1 | 5 | 108 7 1 | 18 | 3 | 1 | 22 | 48 4 1 | 0 | 4 | 1 | 5 | 48 4 1 | 0 | 4 | 1 | 5 | 48 6 1 | 0 | 2 | 1 | 3 |
| 48 6 1 | 0 | 2 | 1 | 3 | 48 7 1 | 0 | 7 | 1 | 8 | 108 2 1 | 0 | 3 | 1 | 4 | 48 6 1 | 0 | 2 | 1 | 3 | 48 6 1 | 0 | 2 | 1 | 3 | 48 7 1 | 0 | 1 | 1 | 2 |
| 60 1 1 | 12 | 3 | 1 | 16 | 15 2 1 | 33 | 3 | 1 | 37 | 108 5 1 | 0 | 3 | 1 | 4 | 60 1 1 | 12 | 3 | 1 | 16 | 60 1 1 | 12 | 3 | 1 | 16 | 60 1 1 | 12 | 2 | 1 | 15 |
| 90 6 1 | 30 | 5 | 1 | 36 | 90 6 1 | 75 | 4 | 1 | 80 | 126 7 1 | 18 | 2 | 1 | 21 | 90 6 1 | 30 | 5 | 1 | 36 | 90 6 1 | 30 | 5 | 1 | 36 | 90 1 1 | 30 | 0 | 1 | 31 |
| 90 1 1 | 0 | 3 | 1 | 4 | 90 1 1 | 0 | 3 | 1 | 4 | 168 1 1 | 42 | 2 | 1 | 45 | 90 1 1 | 0 | 3 | 1 | 4 | 90 1 1 | 0 | 3 | 1 | 4 | 90 4 1 | 0 | 3 | 1 | 4 |
| 90 4 1 | 0 | 3 | 1 | 4 | 90 4 1 | 0 | 3 | 1 | 4 | 168 5 1 | 0 | 4 | 1 | 5 | 90 4 1 | 0 | 3 | 1 | 4 | 90 4 1 | 0 | 3 | 1 | 4 | 90 6 1 | 0 | 2 | 1 | 3 |
| 108 7 1 | 18 | 3 | 1 | 22 | 108 7 1 | 18 | 3 | 1 | 22 | 168 4 1 | 0 | 7 | 1 | 8 | 108 7 1 | 18 | 3 | 1 | 22 | 108 7 1 | 18 | 3 | 1 | 22 | 108 2 1 | 18 | 4 | 1 | 23 |
| 108 2 1 | 0 | 3 | 1 | 4 | 108 2 1 | 0 | 3 | 1 | 4 | 179 4 1 | 11 | 0 | 1 | 12 | 108 2 1 | 0 | 3 | 1 | 4 | 108 2 1 | 0 | 3 | 1 | 4 | 108 5 1 | 0 | 3 | 1 | 4 |
| 108 5 1 | 0 | 3 | 1 | 4 | 108 5 1 | 0 | 3 | 1 | 4 | 179 2 1 | 0 | 6 | 1 | 7 | 108 5 1 | 0 | 3 | 1 | 4 | 108 5 1 | 0 | 3 | 1 | 4 | 108 7 1 | 0 | 2 | 1 | 3 |
| 126 7 1 | 18 | 2 | 1 | 21 | 126 7 1 | 18 | 2 | 1 | 21 | 179 6 1 | 0 | 4 | 1 | 5 | 126 7 1 | 18 | 2 | 1 | 21 | 126 7 1 | 18 | 2 | 1 | 21 | 126 7 1 | 18 | 0 | 1 | 19 |
| 168 1 1 | 42 | 2 | 1 | 45 | 168 1 1 | 42 | 2 | 1 | 45 | 196 7 1 | 17 | 1 | 1 | 19 | 168 1 1 | 42 | 2 | 1 | 45 | 168 1 1 | 42 | 2 | 1 | 45 | 168 1 1 | 42 | 2 | 1 | 45 |
| 168 5 1 | 0 | 4 | 1 | 5 | 168 5 1 | 0 | 4 | 1 | 5 | 60 1 1 | 136 | 2 | 1 | 139 | 168 5 1 | 0 | 4 | 1 | 5 | 168 5 1 | 0 | 4 | 1 | 5 | 168 4 1 | 0 | 3 | 1 | 4 |
| 168 4 1 | 0 | 7 | 1 | 8 | 168 4 1 | 0 | 7 | 1 | 8 | 48 6 1 | 12 | 5 | 1 | 18 | 168 4 1 | 0 | 7 | 1 | 8 | 168 4 1 | 0 | 7 | 1 | 8 | 168 5 1 | 0 | 1 | 1 | 2 |
| 179 4 1 | 11 | 0 | 1 | 12 | 179 4 1 | 11 | 0 | 1 | 12 | 48 4 1 | 0 | 6 | 1 | 7 | 179 4 1 | 11 | 0 | 1 | 12 | 179 4 1 | 11 | 0 | 1 | 12 | 179 2 1 | 11 | 5 | 1 | 17 |
| 179 2 1 | 0 | 6 | 1 | 7 | 179 2 1 | 0 | 6 | 1 | 7 | 48 0 1 | 0 | 4 | 1 | 5 | 179 2 1 | 0 | 6 | 1 | 7 | 179 2 1 | 0 | 6 | 1 | 7 | 179 4 1 | 0 | 2 | 1 | 3 |
| 179 6 1 | 0 | 4 | 1 | 5 | 179 6 1 | 0 | 4 | 1 | 5 | 48 7 1 | 0 | 7 | 1 | 8 | 179 6 1 | 0 | 4 | 1 | 5 | 179 6 1 | 0 | 4 | 1 | 5 | 179 6 1 | 0 | 2 | 1 | 3 |
| 196 7 1 | 17 | 1 | 1 | 19 | 196 7 1 | 17 | 1 | 1 | 19 | 15 2 1 | 33 | 3 | 1 | 37 | 196 7 1 | 17 | 1 | 1 | 19 | 196 7 1 | 17 | 1 | 1 | 19 | 196 7 1 | 17 | 1 | 1 | 19 |
| Total | 231 | 67 | 20 | 318 | Total | 231 | 75 | 20 | 326 | Total | 312 | 70 | 20 | 402 | Total | 231 | 67 | 20 | 318 | Total | 231 | 67 | 20 | 318 | Total | 231 | 51 | 20 | 302 |
| Avg | 11.55 | 3.35 | 1.00 | 15.90 | Avg | 11.55 | 3.75 | 1.00 | 16.30 | Avg | 15.60 | 3.50 | 1.00 | 20.10 | Avg | 11.55 | 3.35 | 1.00 | 15.90 | Avg | 11.55 | 3.35 | 1.00 | 15.90 | Avg | 11.55 | 2.55 | 1.00 | 15.10 |

Figure 4: Working of all Disk-Scheduling algorithms for Case-1

| CASE-1 ( Sp Asc) | | | | | | | | | | | |
|---|---|---|---|---|---|---|---|---|---|---|---|
| | FCFS | SSTF | SCAN | CSCAN | LOOK | C-LOOK | ODSA | HDSA | RP-10 | SMCC | MRSA | MODSBSM |
| TSKT | 231.00 | 231.00 | 261.00 | 373.00 | 231.00 | 337.00 | 231.00 | 231.00 | 312.00 | 231.00 | 231.00 | 231.00 |
| TRL | 67.00 | 75.00 | 75.00 | 90.00 | 75.00 | 90.00 | 67.00 | 75.00 | 70.00 | 67.00 | 72.00 | 51.00 |
| TDTT | 20.00 | 20.00 | 20.00 | 20.00 | 20.00 | 20.00 | 20.00 | 20.00 | 20.00 | 20.00 | 20.00 | 20.00 |
| TDAT | 318.00 | 326.00 | 356.00 | 483.00 | 326.00 | 447.00 | 318.00 | 326.00 | 402.00 | 318.00 | 318.00 | 302.00 |
| ADAT | 15.90 | 16.30 | 17.80 | 24.15 | 16.30 | 22.35 | 15.90 | 16.30 | 20.10 | 15.90 | 15.90 | 15.10 |

Figure 5: Performance measure of all disk-scheduling algorithms for case-1

### 4.3 Case-2

Here we considered a situation with all memory request in descending order in Single Platter Hard-disk Figure 6 represents the memory request of Case-2. Initially the head is positioned at 75t1p7s index. Figure 7 represents the working of traditional, referred, and proposed disk scheduling algorithm to access all Memory Requests of Case-2. Figure 8 represents the total seek time and disk access time required with respect to each disk scheduling algorithms for complete all memory requests as per Case-2.

Figure 6: Disk queue for memory request of case-2

Figure 7: Working of all Disk-Scheduling algorithms for Case-2

| | FCFS | SSTF | SCAN | CSCAN | LOOK | C-LOOK | ODSA | HDSA | RP-10 | SMCC | MRSA | MODSBSM |
|---|---|---|---|---|---|---|---|---|---|---|---|---|
| | | | | | | CASE-2 ( Sp Desc) | | | | | | |
| TSKT | 267.00 | 204.00 | 260.00 | 367.00 | 204.00 | 283.00 | 204.00 | 204.00 | 267.00 | 204.00 | 204.00 | 204.00 |
| TRL | 80.00 | 78.00 | 78.00 | 77.00 | 78.00 | 77.00 | 70.00 | 78.00 | 80.00 | 70.00 | 70.00 | 62.00 |
| TDTT | 20.00 | 20.00 | 20.00 | 20.00 | 20.00 | 20.00 | 20.00 | 20.00 | 20.00 | 20.00 | 20.00 | 20.00 |
| TDAT | 367.00 | 302.00 | 358.00 | 464.00 | 302.00 | 380.00 | 294.00 | 302.00 | 367.00 | 294.00 | 294.00 | 286.00 |
| ADAT | 18.35 | 15.10 | 17.90 | 23.20 | 15.10 | 19.00 | 14.70 | 15.10 | 18.35 | 14.70 | 14.70 | 14.30 |

Figure 8: Performance measure of all disk-scheduling algorithms for case-2

## 4.4 Case-3

Here we considered a situation with all memory request in random order in Single Platter Harddisk. Figure 9 represents the memory request of Case-3. Initially the head is positioned at 165t1p7s index. Figure 10 represents the working of traditional, referred, and proposed disk scheduling algorithm to access all Memory Requests of Case-3. Figure 11 represents the total seek time and disk access time required with respect to each disk scheduling algorithms for complete all memory requests as per Case-3.

| Platter | 1 | 1 | 1 | 1 | 1 | 1 | 1 | 1 | 1 | 1 | 1 | 1 | 1 | 1 | 1 | 1 | 1 | 1 |
|---|---|---|---|---|---|---|---|---|---|---|---|---|---|---|---|---|---|---|
| Track | 45 | 98 | 15 | 98 | 160 | 198 | 15 | 45 | 98 | 160 | 198 | 65 | 45 | 160 | 198 | 113 | 15 | 59 | 15 | 5 |
| Sector | 5 | 2 | 3 | 6 | 2 | 1 | 0 | 0 | 4 | 7 | 6 | 2 | 6 | 6 | 4 | 4 | 6 | 0 | 4 | 2 |
| Index | 45t1p5s | 98t1p2s | 15t1p3s | 98t1p6s | 160t1p2s | 198t1p1s | 15t1p0s | 45t1p0s | 98t1p4s | 160t1p7s | 198t1p6s | 65t1p2s | 45t1p6s | 160t1p6s | 198t1p4s | 113t1p4s | 15t1p6s | 59t1p0s | 15t1p4s | 5t1p2s |
| R/w | | | | | | | | | | | | | | | | | | | | |
| BSI | 0 | 0 | 0 | 0 | 0 | 0 | 0 | 0 | 0 | 0 | 0 | 0 | 0 | 0 | 0 | 0 | 0 | 0 | 0 | 0 |

Figure 9: Disk queue for memory request of case-3

Figure10: Working of all Disk-Scheduling algorithms for Case-3

International Journal of Computer Science, Engineering and Applications (IJCSEA) Vol. 9, No. 1/2/3, June 2019

| CASE-3 ( Sp Rand) | | | | | | | | | | | |
|---|---|---|---|---|---|---|---|---|---|---|---|
| | FCFS | SSTF | SCAN | CSCAN | LOOK | C-LOOK | ODSA | HDSA | RP-10 | SMCC | MRSA | MODSBSM |
| TSKT | 1,392.00 | 236.00 | 363.00 | 365.00 | 353.00 | 353.00 | 226.00 | 226.00 | 226.00 | 226.00 | 226.00 | 226.00 |
| TRL | 75.00 | 75.00 | 77.00 | 66.00 | 77.00 | 66.00 | 67.00 | 67.00 | 67.00 | 67.00 | 67.00 | 59.00 |
| TDTT | 20.00 | 20.00 | 20.00 | 20.00 | 20.00 | 20.00 | 20.00 | 20.00 | 20.00 | 20.00 | 20.00 | 20.00 |
| TDAT | 1,487.00 | 331.00 | 460.00 | 451.00 | 450.00 | 439.00 | 313.00 | 313.00 | 313.00 | 313.00 | 313.00 | 305.00 |
| ADAT | 74.35 | 16.55 | 23.00 | 22.55 | 21.95 | 21.95 | 15.65 | 15.65 | 15.65 | 15.65 | 15.65 | 15.25 |

Figure 11: Performance measure of all disk-scheduling algorithms for case

### 4.5 Case-4

Here we considered a situation with all memory request in ascending order in Multi Platter Hard-disk. Figure 12 represents the memory request of Case-4. Initially the head is positioned at 140t1p0s index. Figure 13 represents the working of traditional, referred, and proposed disk scheduling algorithm to access all Memory Requests of Case-4. Figure 14 represents the total seek time and disk access time required with respect to each disk scheduling algorithms for complete all memory requests as per Case-4.

| Platter | 1 | 2 | 2 | 3 | 2 | 1 | 4 | 1 | 3 | 4 | 2 | 1 | 2 | 4 | 1 | 3 | 3 | 3 | 2 | 2 |
|---|---|---|---|---|---|---|---|---|---|---|---|---|---|---|---|---|---|---|---|---|
| Track | 18 | 25 | 25 | 25 | 32 | 46 | 46 | 46 | 78 | 95 | 95 | 95 | 95 | 123 | 148 | 156 | 156 | 156 | 156 | 197 |
| Sector | 3 | 6 | 4 | 0 | 2 | 4 | 7 | 2 | 5 | 2 | 5 | 1 | 6 | 2 | 0 | 6 | 7 | 5 | 1 | 7 |
| Index | 18t1p3s | 25t2p6s | 25t2p4s | 25t3p0s | 32t2p2s | 46t1p4s | 46t4p7s | 46t1p2s | 78t3p5s | 95t4p2s | 95t2p5s | 95t1p1s | 95t2p6s | 123t4p2s | 148t1p0s | 156t3p6s | 156t3p7s | 156t3p5s | 156t2p1s | 197t2p7s |
| R/w | | | | | | | | | | | | | | | | | | | | |
| BSI | 0 | 0 | 0 | 0 | 0 | 0 | 0 | 0 | 0 | 0 | 0 | 0 | 0 | 0 | 0 | 0 | 0 | 0 | 0 | 0 |

Figure 12: Disk queue for memory request of case-4

| | FCFS | | | | | SSTF | | | | | SCAN | | | | | CSCAN | | | | | LOOK | | | | | CLOOK | | | | |
|---|---|---|---|---|---|---|---|---|---|---|---|---|---|---|---|---|---|---|---|---|---|---|---|---|---|---|---|---|---|---|
| T S P | ST | RL | DTT | DAT | T S P | ST | RL | DTT | DAT | T S P | ST | RL | DTT | DAT | T S P | ST | RL | DTT | DAT | T S P | ST | RL | DTT | DAT | T S P | ST | RL | DTT | DAT | | |

Figure 13: Working of all Disk-Scheduling algorithms for Case-4

| CASE-4 ( Mp Asc) | | | | | | | | | | | | |
|---|---|---|---|---|---|---|---|---|---|---|---|---|
| | FCFS | SSTF | SCAN | CSCAN | LOOK | C-LOOK | ODSA | HDSA | RP-10 | SMCC | MRSA | MODSBSM |
| TSKT | 301.00 | 333.00 | 240.00 | 381.00 | 236.00 | 341.00 | 236.00 | 236.00 | 236.00 | 236.00 | 236.00 | 236.00 |
| TRL | 79.00 | 79.00 | 83.00 | 74.00 | 83.00 | 74.00 | 83.00 | 83.00 | 83.00 | 83.00 | 83.00 | 67.00 |
| TDTT | 45.00 | 45.00 | 44.00 | 43.00 | 44.00 | 43.00 | 46.00 | 44.00 | 44.00 | 46.00 | 46.00 | 42.00 |
| TDAT | 425.00 | 457.00 | 367.00 | 498.00 | 363.00 | 458.00 | 365.00 | 363.00 | 363.00 | 365.00 | 365.00 | 345.00 |
| ADAT | 21.25 | 22.85 | 18.35 | 24.90 | 18.15 | 22.90 | 18.25 | 18.15 | 18.15 | 18.25 | 18.25 | 17.25 |

Figure14: Performance measure of all disk-scheduling algorithms for case-4

### 4.6  Case-5

Here we considered a situation with all memory request in descending order in Multi Platter Hard-disk. Figure 15 represents the memory request of Case-5. Initially the head is positioned at 65t1p4s index. Figure 16 represents the working of traditional, referred, and proposed disk scheduling algorithm to access all Memory Requests of Case-5. Figure 17 represents the total seek time and disk access time required with respect to each disk scheduling algorithms for complete all memory requests as per Case-5.

| Platter | 4 | 1 | 2 | 1 | 2 | 4 | 2 | 3 | 4 | 4 | 4 | 2 | 1 | 1 | 4 | 1 | 3 | 4 | 1 | 2 |
|---|---|---|---|---|---|---|---|---|---|---|---|---|---|---|---|---|---|---|---|---|
| Track | 196 | 167 | 167 | 167 | 167 | 143 | 143 | 126 | 126 | 126 | 98 | 98 | 98 | 63 | 63 | 63 | 42 | 19 | 19 | 19 |
| Sector | 7 | 4 | 7 | 2 | 3 | 2 | 0 | 1 | 6 | 4 | 6 | 5 | 2 | 7 | 0 | 6 | 4 | 6 | 2 | 4 |
| Index | 196t4p7s | 167t1p4s | 167t2p7s | 167t1p2s | 167t2p3s | 143t4p2s | 143t2p0s | 126t3p1s | 126t4p6s | 126t4p4s | 98t4p6s | 98t2p5s | 98t1p2s | 63t1p7s | 63t4p0s | 63t1p6s | 42t3p4s | 19t4p6s | 19t1p2s | 19t2p4s |
| R/w | | | | | | | | | | | | | | | | | | | | |
| BSI | 0 | 0 | 0 | 0 | 0 | 0 | 0 | 0 | 0 | 0 | 0 | 0 | 0 | 0 | 0 | 0 | 0 | 0 | 0 | 0 |

Figure 15: Disk queue for memory request of case-5

Figure 2: Working of all Disk-Scheduling algorithms for Case-5

| CASE-5 ( Mp Desc) | | | | | | | | | | | | |
|---|---|---|---|---|---|---|---|---|---|---|---|---|
| | FCFS | SSTF | SCAN | CSCAN | LOOK | C-LOOK | ODSA | HDSA | RP-10 | SMCC | MRSA | MODSBSM |
| TSKT | 308.00 | 223.00 | 314.00 | 396.00 | 308.00 | 352.00 | 223.00 | 223.00 | 308.00 | 223.00 | 223.00 | 223.00 |
| TRL | 80.00 | 75.00 | 72.00 | 83.00 | 72.00 | 83.00 | 75.00 | 75.00 | 72.00 | 75.00 | 75.00 | 51.00 |
| TDTT | 51.00 | 49.00 | 51.00 | 50.00 | 51.00 | 50.00 | 49.00 | 49.00 | 51.00 | 49.00 | 49.00 | 45.00 |
| TDAT | 439.00 | 347.00 | 437.00 | 529.00 | 431.00 | 485.00 | 347.00 | 347.00 | 431.00 | 347.00 | 347.00 | 319.00 |
| ADAT | 21.95 | 17.35 | 21.85 | 26.45 | 21.55 | 24.25 | 17.35 | 17.35 | 21.55 | 17.35 | 17.35 | 15.95 |

Figure 17: Performance measure of all disk-scheduling algorithms for case-5

### 4.7 Case-6

Here we considered a situation with all memory request in random order in Multi Platter Hard-disk. Figure 18 represents the memory request of Case-6. Initially the head is positioned at 90t1p6s. Figure 19 represents the working of traditional, referred, and proposed disk scheduling algorithm to access all Memory Requests of Case-6. Figure 20 represents the total seek time and disk access time required with respect to each disk scheduling algorithms for complete all memory requests as per Case-6.

| Platter | 3 | 4 | 2 | 4 | 1 | 2 | 4 | 4 | 4 | 4 | 3 | 2 | 3 | 2 | 1 | 2 | 2 | 4 | 2 | 1 |
|---|---|---|---|---|---|---|---|---|---|---|---|---|---|---|---|---|---|---|---|---|
| Track | 55 | 15 | 105 | 48 | 83 | 165 | 39 | 83 | 22 | 165 | 105 | 15 | 83 | 165 | 39 | 15 | 105 | 39 | 22 | 165 |
| Sector | 1 | 6 | 4 | 2 | 6 | 3 | 6 | 7 | 7 | 5 | 1 | 4 | 3 | 7 | 2 | 0 | 6 | 7 | 0 | 6 |
| Index | 55t3p1s | 15t4p6s | 105t2p4s | 48t4p2s | 83t1p6s | 165t2p3s | 39t4p6s | 83t4p7s | 22t4p7s | 165t4p5s | 105t3p1s | 15t2p4s | 83t3p3s | 165t2p7s | 39t1p2s | 15t2p0s | 105t2p6s | 39t4p7s | 22t2p0s | 165t1p6s |
| R/w | | | | | | | | | | | | | | | | | | | | |
| BSI | 0 | 0 | 0 | 0 | 0 | 0 | 0 | 0 | 0 | 0 | 0 | 0 | 0 | 0 | 0 | 0 | 0 | 0 | 0 | 0 |

Figure 18: Disk queue for memory request of case-6

Figure 19: Working of all Disk-Scheduling algorithms for Case-6

| CASE-6 ( Mp Rand) | | | | | | | | | | | |
|---|---|---|---|---|---|---|---|---|---|---|---|
| | FCFS | SSTF | SCAN | CSCAN | LOOK | C-LOOK | ODSA | HDSA | RP-10 | SMCC | MRSA | MODSBSM |
| TSKT | 1,479.00 | 269.00 | 293.00 | 391.00 | 225.00 | 293.00 | 225.00 | 225.00 | 225.00 | 225.00 | 225.00 | 225.00 |
| TRL | 80.00 | 80.00 | 80.00 | 85.00 | 80.00 | 85.00 | 72.00 | 80.00 | 80.00 | 72.00 | 72.00 | 57.00 |
| TDTT | 44.00 | 52.00 | 51.00 | 52.00 | 51.00 | 52.00 | 52.00 | 52.00 | 51.00 | 52.00 | 49.00 | 47.00 |
| TDAT | 1,603.00 | 401.00 | 424.00 | 528.00 | 356.00 | 430.00 | 349.00 | 357.00 | 356.00 | 349.00 | 347.00 | 329.00 |
| ADAT | 80.15 | 20.05 | 21.20 | 26.40 | 17.80 | 21.50 | 17.45 | 17.85 | 17.80 | 17.45 | 17.35 | 16.45 |

Figure 20: Performance measure of all disk-scheduling algorithms for case-6

## 4.8 Result Analysis

From the analysis of all the six cases, it is concluded that our algorithm performs better than all traditional and latest modified disk scheduling algorithms, as per prescribed performance measure of the disk scheduling algorithm. Table 3 represents the total performance measures of all algorithms, considering all six cases. All five derived algorithms (ODSA, HDSA, RP-10, SMCC, MRSA) focused on minimising the seek time. They did not consider to minimize the rotational latency or data transfer time. The proposed disk scheduling algorithm minimises all three performance measure of disk scheduling algorithm, Viz. seek time, rotational latency, data transfer time. Therefore, the proposed algorithm was able to optimize the total disk access time by 33.53% and 7.51% in comparison with traditional and referred disk scheduling algorithms respectively. Figure 21 represents the comparative analysis of average disk access time of all algorithms in each case.

Along with that our algorithm will also be able to detect and manage the bad sectors of the hard-disk. Once a memory index detected twice as inaccessible memory, then it transfers that index to "Bad Sector List" table. Then it resolves the situation by finding a proper prescribed bit for that index. This reduces n-2 number of un-necessary memory read, which in turn reduces the power consumption of hard-disk by:
Energy Conserved = e*n-2. (e ~ 100 fj => Energy Consumed per one-bit memory access)

Table3 : Total performance measures of all algorithms

|  | TOTAL | | | | | | | | | | | |
| --- | --- | --- | --- | --- | --- | --- | --- | --- | --- | --- | --- | --- |
|  | FCFS | SSTF | SCAN | CSCAN | LOOK | CLOOK | ODSA | HDSA | RP-10 | SMCC | MRSA | MODSBSM |
| TSKT | 3978 | 1496 | 1731 | 2273 | 1557 | 1959 | 1345 | 1345 | 1574 | 1345 | 1345 | 1345 |
| TRL | 461 | 462 | 465 | 475 | 465 | 475 | 434 | 458 | 452 | 434 | 439 | 347 |
| TDTT | 200 | 206 | 206 | 205 | 206 | 205 | 207 | 205 | 206 | 207 | 204 | 194 |
| TDAT | 4639 | 2164 | 2402 | 2953 | 2228 | 2639 | 1986 | 2008 | 2232 | 1986 | 1984 | 1886 |
| ADAT | 38.66 | 18.03 | 20.02 | 24.61 | 18.48 | 21.99 | 16.55 | 16.73 | 18.60 | 16.55 | 16.53 | 15.72 |

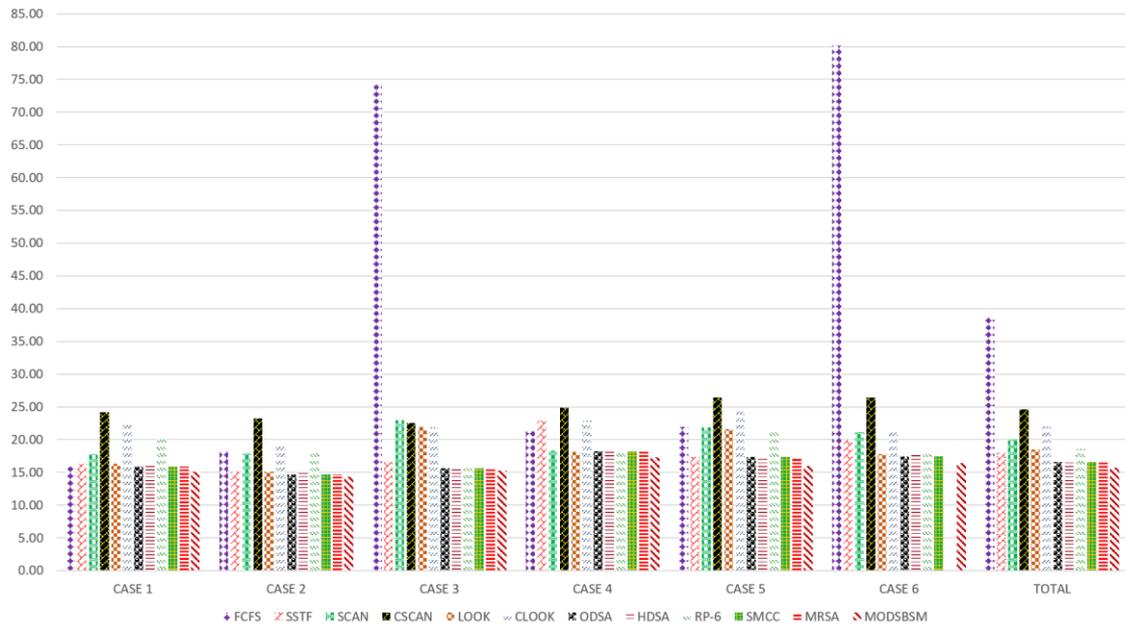

Figure 21: Comparative analysis of average disk access time of all algorithms

As this algorithm resolve bad sector, the read/write head does not need to access a single index (a particular position in hard-disk), multiple times, with the help of infrared/ blue-ray beams. So this algorithm also reduces heat generation within the hard-disk by:
Heat reduced = h*n-2 (heat generated per bit access).

## 5. CONCLUSIONS

In this paper we proposed a new disk scheduling algorithm, MODSBSM. Here we demonstrate the comparative analysis of proposed algorithm with respect to all traditional disk scheduling algorithm, has been carried out in both single-platter and multi-Platter Hard-disk. The proposed algorithm provides better performance metrics by minimizing the average disk access time. This algorithm also able to detect and resolve the bad sectors of hard-disk. In future we want to propose another scheduling algorithm for implementing in SSD.

## Authors

**Amar Ranjan Dash** obtained his B. Tech Comp. Sc. from Biju Patnaik University of Technology, India and M. Tech degree from Berhampur University, India. Currently, He is pursuing his doctoral research. He has Ten research publications to his credit. His research interests include Web Accessibility, Algorithm Optimization, and cloud computing. He is a member of ACM & IEEE.

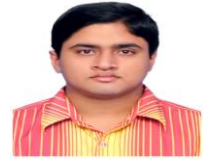

**Sandipta Kumar Sahu** has achieved his B. Tech. Comp. Sc. and M. Tech. Comp. Sc. from Biju Patnaik University of Technology, India. He has two research publication to his credit. His research interests include Computer Architecture, Compiler Design, and Operating System.

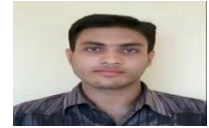

**B. Kewal** has achieved his BCA degree from Berhampur University, Odisha, India. Currently, He is pursuing his MCA Post graduation from Berhampur University, Odisha, India. His research interests include Operating System, Algorithm optimization, and Compiler Design.

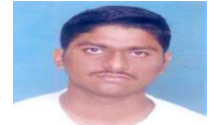